\documentclass[twocolumn,jcp,showpacs]{revtex4}

\usepackage{psfrag}
\usepackage{graphicx}
\usepackage{amsmath}
\usepackage{amssymb}

\begin{document}
\title{Benchmark density functional theory calculations for nano-scale
conductance.}
\author{M. Strange}
\affiliation{Center for Atomic-scale Materials Design, Department of
Physics \\ Technical University of Denmark, DK - 2800 Kgs. Lyngby, Denmark}
\author{I. S. Kristensen}
\affiliation{Center for Atomic-scale Materials Design, Department of
Physics \\ Technical University of Denmark, DK - 2800 Kgs. Lyngby, Denmark}
\author{K. S. Thygesen}
\affiliation{Center for Atomic-scale Materials Design, Department of
Physics \\ Technical University of Denmark, DK - 2800 Kgs. Lyngby, Denmark}

\date{\today}

\begin{abstract}
We present a set of benchmark calculations for the Kohn-Sham elastic transmission function of five representative single-molecule junctions. The transmission functions are calculated using two different density functional theory (DFT) methods, namely an ultrasoft pseudopotential plane wave code in combination with maximally localized Wannier functions, and the norm-conserving pseudopotential code Siesta which applies an atomic orbital basis set. For all systems we find that the Siesta transmission functions converge toward the plane-wave result as the Siesta basis is enlarged. Overall, we find that an atomic basis with double-zeta and polarization is sufficient (and in some cases even necessary) to ensure quantitative agreement with the plane-wave calculation. We observe a systematic down shift of the Siesta transmission functions relative to the plane-wave results. The effect diminishes as the atomic orbital basis is enlarged, however, the convergence can be rather slow.  

\end{abstract}
\pacs{72.10-d,73.40.Cg,73.63.Rt}
\maketitle

\begin{section}{Introduction}\label{sec:intro}
  First-principles calculations of electrical conductance in nano-scale
  contacts represents a main challenge in computational nanophysics. The interest for this type of calculations began in the
  mid-nineties where advances in
  experimental techniques made it possible to contact individual 
  molecules thereby making it possible study the transport of electrons through true
  nano-scale structures~\cite{agrait_report,nitzan03}. Apart from the
  scientific interest, the development of reliable simulation tools
  for nano-scale quantum transport is relevant in
  relation to the continued miniaturization of conventional semi-conductor
  electronics, but also for the introduction of the new generation of
  molecule based electronics.

  It has by now become standard to calculate conductance in nano-scale
  contacts by employing a combination of non-equilibrium Green's
  function theory (NEGF) and ground state density functional theory
  (DFT). The resulting NEGF-DFT formalism provides a numerically efficient way
  of evaluating the Landauer-B{\"u}ttiker conductance due to electrons
  moving in the effective Kohn-Sham (KS) potential without having to
  calculate the scattering states explicitly. It has been applied
  extensively to a number of different systems
  ranging from pure metallic contacts, over organic molecules to
  carbon nanotubes suspended between metallic electrodes. Overall the
  approach has been successful in describing qualitative features and
  trends~\cite{cuevas,cms_stokbro_2003}, however, quantitative agreement with
experiments has mainly
  been obtained for strongly coupled systems such as metallic point
  contacts, monatomic chains, as well as junctions containing small
  chemisorbed molecules~\cite{prl_pt_osc,prl_thygesen_h2,prb_strange_pt_co}.

  It is generally accepted that the NEGF-DFT method only provides an
  approximation to the true conductance - even if the exact
  exchange-correlation (xc-)functional could be used, and the quality of the result is expected to be strongly system dependent. Moreover, it is not easy to estimate the effect
  of using approximate xc-functionals such as the LDA or GGA. We
  mention here that more sophisticated methods for quantum transport based on
  configuration interaction, the $GW$ method, time-dependent DFT, and
  the Kubo formula have recently been
proposed~\cite{delaney,darancet,thygesen_gw,kurth,bokes}.
  However, these schemes are considerably more demanding than the
  NEGF-DFT and at present they cannot replace NEGF-DFT in practical applications.

  Irrespective of the validity of the NEGF-DFT approach and the role
  played by the approximate functionals, it remains important to
  establish a general consensus concerning the exact result of a
  NEGF-DFT calculation for a given xc-functional and specified system geometry, i.e. a
  benchmark. Although this might seem trivial, the present situation
  is rather unsatisfactory as different results have been published by
  different groups for the same or very similar systems (several
  examples will be given in the text). Perhaps the best example is provided by benzene
  di-thiolate between gold contacts where the calculated conductance
  vary with up to a two orders of magnitude for similar geometries~\cite{prl_pantelides_2007,prl_lang_2000,cms_stokbro_2003,ratner_2006,prb_faleev_2005,jcp_sankey_2004,prl_sanvito_asic_2007}.

  The relatively large variation of the results indicates that the
  conductance, or more generally the elastic transmission function, is
  a highly sensitive quantity. Indeed, the implementation of the open boundary
conditions
  defining the scattering problem represents some numerical challenges.
  Indeed, small errors in the
  description of the coupling between the finite scattering region and
  the infinite leads as well as improper $k$-point samplings in supercell approaches, can introduce significant errors in the resulting
  transmission function.

  In this paper we take a first step towards establishing a common
  reference for NEGF-DFT calculations by performing benchmark
  calculations for a set of five representative nano-scale contacts.
  The benchmarking is achieved by comparing the transmission
  function obtained using two different and independent, albeit
  similar, NEGF-DFT methods: In one case the Hamiltonian is obtained from the
  Siesta DFT program which uses a basis of localized pseudo atomic orbitals 
  (PAOs) together with norm conserving pseudopotentials. The second method
  applies a basis of maximally localized Wannier functions (WFs)
  obtained from the Dacapo DFT code which uses plane-waves and ultra
  soft pseudopotentials. In both cases we use periodic boundary
  conditions in the directions perpendicular to the transport
  direction and we apply the PBE xc-functional \cite{PBE}.

  The five reference systems we have chosen for our benchmark study
  are: (i) A monatomic gold chain with a single CO molecule adsorbed.
  (ii) A 3-atom Pt chain suspended between Pt electrodes.  (iii) An
  $\text{H}_2$ molecule bridging two Pt electrodes. (iv) Benzene-dithiolate (BDT) between Au electrodes, and (v) Bipyridine between Au electrodes. The systems have been chosen
  according to the criterion that both experimental data as well as previous
  NEGF-DFT calculations are available in the literature. Furthermore they
  are representative in the sense that they cover a broad
  class of systems: homogeneous and heterogeneous,
  computationally simple (one-dimensional) and more complex, and strongly as well as
weakly coupled.

 \begin{figure}[!h]
   \includegraphics[width=0.95\linewidth,angle=0]{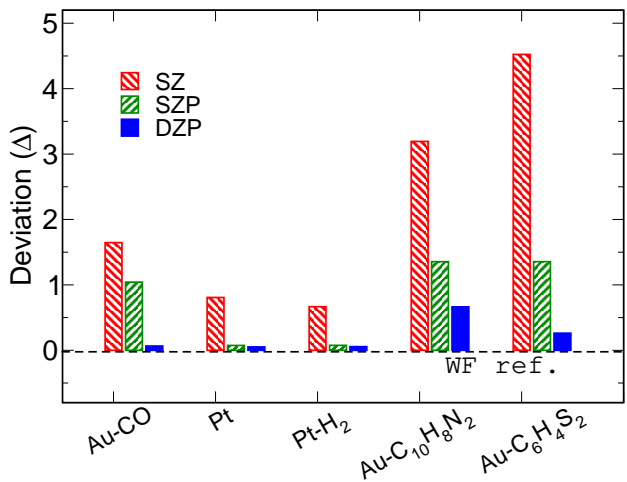}
   \caption[cap.Pt6]{\label{fig.fig0} (color online).
                    Deviation between the WF and Siesta transmission functions for for the five reference systems studied.                       
               The dashed line indicate zero deviation from the WF transmission. Notice that the Siesta results converge toward the WF result as the PAO basis is enlarged.
                    }
  \end{figure}

A main results of our work is summarized in Fig. \ref{fig.fig0} where we show the overall deviation
\begin{equation}\label{eq:deviation}  \Delta=\int_{-\infty}^{\varepsilon_F+E_0}|T_\text{WF}(\varepsilon)-T_\text{PAO}(\varepsilon)|\text{d}\varepsilon,
  \end{equation}
between the transmission functions calculated using the WF and PAO basis sets, respectively. The cutoff energy $E_0$ is taken to be the energy above which the WFs are no longer able to reproduce
  the exact KS eigenstates of the system which is typically $\sim 3$~eV above the Fermi level. For all the systems we find
  that the deviation $\Delta$ decreases as the Siesta basis is enlarged meaning that the Siesta transmission functions
  converge toward the WF result. We take this as evidence for the correctness of 
the WF results and the justification for the use of the term \emph{benchmark} calculation.
  
  In general we find that the double-zeta polarized (DZP)
  basis provides very good agreement with the WF basis, whereas the single-zeta polarized (SZP) and, in particular, the single-zeta (SZ) basis can produce substantially incorrect features in the transmission function.

The paper is organized as follows. In Sec.~\ref{sec:method} we briefly
review the NEGF-DFT formalism and introduce the two specific
implementations used in the present study. In Secs.~\ref{sec:AuCO}-\ref{sec:AuBDT} we
present the benchmark calculations for the five reference systems and in 
Sec.~\ref{sec:conclusions} we give our conclusions.

\end{section}

\begin{section}{Method}\label{sec:method}
  In this section we first outline the NEGF-DFT method which has
  become standard for nanoscale conductance calculations. The two
  specific NEGF-DFT implementations applied in the present work are
  then introduced and their key parameters are discussed. We then consider the
important issue, common
  to both methods, of how to treat periodic boundary conditions in the
  plane perpendicular to the transport direction. We end the section
  with a discussion of the advantages and disadvantages of the two
  methods.

\begin{subsection}{NEGF-DFT}
  The zero temperature, linear response conductance due to non-interacting electrons
  scattering off a central region ($C$) connected to thermal
  reservoirs via two ballistic leads ($L$ and $R$), can be written as
\begin{equation}
G=G_0T(\varepsilon_F),
\end{equation}
where $T(\varepsilon)$ is the elastic transmission function and
$G_0=2e^2/h$ is the quantum unit of conductance. Using the NEGF
formalism, Meir and Wingreen have derived a very useful formula which
expresses the transmission function in terms of the Green's function
of the central region~\cite{meir92},
\begin{equation}\label{eq.trans}
T(\varepsilon)=\mathrm{Tr}[G^r(\varepsilon)\Gamma_L(\varepsilon)G^a(\varepsilon)\Gamma_R(\varepsilon)].
\end{equation}
In this expression the trace runs over the central region basis
functions and $\Gamma_{L/R}$ is obtained from the lead self-energies
(defined in Eq.~(\ref{eq.sigma}) below) as
$\Gamma_{L/R}=i(\Sigma_{L/R}-\Sigma_{L/R}^{\dagger})$.

In the NEGF-DFT method both the leads and central region are
modeled by the effective KS Hamiltonian, $\hat
h_{\text{KS}}=-\frac{1}{2}\nabla^2+v_{\text{eff}}(\mathbf r)$. The
self-consistent effective potential consists of the well known parts
$v_{\text{eff}}=v_{\text{ext}}+v_H+v_{xc}$. Introducing a basis of localized
orbitals, \{$\phi_i$\}, we define the Hamiltonian and overlap matrices by
$H_{ij}=\langle \phi_i|\hat h_{\text{KS}}|\phi_j\rangle$ and $S_{ij}=\langle
\phi_i|\phi_j\rangle$, respectively. In the original derivation by
Meir and Wingreen the basis was assumed orthogonal, but the
generalization to non-orthogonal basis
sets shows that Eq.~(\ref{eq.trans}) still
holds when the Green's function is defined as~\cite{prb_thygesen_nonortho}
\begin{equation}
G(z) = [zS_C-H_C-\Sigma_L(z)-\Sigma_R(z)]^{-1}.
\end{equation}
Here the matrices $H_C$ and $S_C$ are the blocks of $H$ and $S$
corresponding to the central region basis functions. The retarded
Green's function, $G^r(\varepsilon)$, is obtained for
$z=\varepsilon+i0^{+}$, and the advanced Green's function for $z=\varepsilon-i0^+$ or $G^a=(G^r)^\dagger$.

The self-energy of lead $\alpha$ is given by
\begin{equation}\label{eq.sigma}
  \Sigma_{\alpha}(z) =
  (zS_{C \alpha}-H_{C \alpha})g_{\alpha}^0(z)(zS_{\alpha C}-H_{\alpha C}),
\end{equation}
where $H_{C \alpha}$ and $S_{C \alpha}$ are the coupling- and overlap
matrix between basis functions in the central region and lead
$\alpha$, respectively.  $g_{\alpha}^0$ is the surface Green's
function describing the isolated semi-infinite lead,
$g^0_{\alpha}(z)=[zS_{\alpha}-H_{\alpha}]^{-1}$, which can be
calculated recursively using the decimation technique~\cite{guinea}.

\end{subsection}
\begin{subsection}{Method 1: Wannier functions from plane-wave DFT}
  In method 1 the Kohn-Sham Hamiltonian is obtained from an accurate
  plane-wave pseudopotential DFT code~\cite{dacapo}. The ion cores are
  replaced by ultrasoft pseudopotentials~\cite{ultrasoft} and we use an
  energy cutoff of 25 Ry for the plane wave expansion. The Kohn-Sham
  eigenstates are transformed into partly occupied Wannier functions
  (WFs)~\cite{prlprb_thygesen_wannier} which are used to obtain a
  tight-binding like representation of the Hamiltonian. The WFs are constructed such
that
  any eigenstate below a selected energy, $E_0$, can be exactly
  represented by a linear combination of WFs. In the applications we
  have chosen $E_0$ in the range of 2-4~eV above the Fermi level. In this way the
accuracy
  of the plane-wave calculation is carried over to the WF basis for
  all energies relevant for transport. 

  By performing separate DFT calculations for the (periodic) leads and
  $C$ we obtain a set of WFs for each region. Note, that $C$ always contains a few buffer layers of the lead material on both
  sides of the nano-contact to ensure that the KS potential at the
  end planes of $C$ has
  converged to its value in the leads. Since the
  WFs in the lead in general will differ from those in the outermost
  lead unit cells of the central region, care must be taken
  to evaluate the coupling and overlap matrices $H_{C\alpha}$ and
  $S_{S\alpha}$. Notice also that although the WFs by construction are
  orthogonal within each region, WFs
  belonging to different regions will in general be non-orthogonal.
  For more details on the construction of the WFs and the calculations
  of the Hamiltonian matrix for the combined $L-C-R$ system we refer to 
  Ref.~\onlinecite{jcp_thygesen_transport}. We shall refer to the results
  obtained from method 1 as the WF results.
  \end{subsection}

  \begin{subsection}{Method 2: PAO Siesta basis}
    Method 2 is based on the DFT code Siesta~\cite{siesta} which uses
    finite range pseudoatomic orbitals (PAO)~\cite{paos} as basis
    functions and Troullier-Martins norm conserving
    pseudopotentials~\cite{TM}. As in method 1, the Hamiltonians for the
    leads and the central region are obtained from separate
    calculations. Because the KS potential to the left and right of $C$,
    by definition has converged to the value in the leads, we can take
    the coupling between central region and lead $\alpha$,
    $H_{C\alpha}$, from the pure lead calculation. Note that this is
    in contrast to method 1, where the different shape of the WFs in
    the periodic lead and the lead part of the central region makes
    it essential to evaluate the coupling matrix directly. Note also
    that this approximation, i.e. the use of the intra-lead coupling
    matrix elements ($H_{\alpha \alpha}$) in $H_{C\alpha}$, can be
    controlled by including a larger portion of the lead in $C$. In
    practice we find that 3-4 atomic layers must be included in $C$ on
    both sides of the junction to obtain converged conductances.

    In the present study we restrict ourself to the standard PAOs
    for Siesta: SZ, SZP and DZP. For the confinement energy,
    determining the range of PAOs, we use $0.01~\text{Ry}$ and for the
    meshcutoff we use $200~\text{Ry}$.
  \end{subsection}

  \begin{subsection}{Common ingredients}
    In both methods 1 and 2 we treat exchange and correlation
    effects with the PBE energy functional~\cite{PBE}. Furthermore, we
    impose periodic boundary conditions in the surface plane
    directions. This means that we are in fact considering the
    conductance of a periodic array of junctions instead of just a
    single junction. Instead of the localized basis function
    $\phi_n(\mathbf r)$ (this could be a WF or a PAO) we thus consider
    the Bloch function
\begin{equation}
  \chi_{n{\bf k}_\parallel}=\sum_{{\bf R}_\parallel}e^{i{\bf k}_\parallel\cdot
    {\bf R}_\parallel}\phi_n(\bf r-{\bf R}_\parallel),
\end{equation}
where ${\bf R}_\parallel$ runs over supercells in the surface plane
and $\bf k_{\parallel}$ is a wave-vector in the corresponding
two-dimensional Brillouin zone (BZ). Since $\bf k_\parallel$ is a good
quantum number, we can construct the Hamiltonian, $H({\bf
  k}_\parallel)$, for each $\bf k$-point separately. This in turn
implies that the conductance \emph{per} junction is given by the
average
\begin{equation}
  G=\sum_{{\bf k}_{\parallel}}w({\bf k}_{\parallel})G({\bf
    k}_\parallel),
\end{equation}
where $w({\bf k}_\parallel)$ are symmetry determined weight factors.
Unless stated otherwise, we have used a $4\times4$ Monkhorst-Pack
$\mathbf k_\parallel$-point
sampling of the surface BZ, which for all the systems studied yields
conductances converged to within a few
percent~\cite{prb_thygesen_kpoints,jcp_thygesen_transport}.
We take the Fermi level of the bulk lead as the common Fermi level
of the combined system by shifting the levels in the central region by a 
constant. This is done by adding to $H_C$ the matrix $\delta S_c$,
where $\delta = [H_L]_{0,0}-[H_C]_{0,0}$ and the $(0,0)$ element 
corresponds to the onsite energy of a basis function 
located near the interface between $L$ and $C$.

The main advantages of method 1 are: (i) The accuracy of the plane
wave calculation carries over to the WF basis set. (ii) The WFs basis
set is truly minimal and often results in even fewer basis functions
than a SZ basis. The WF basis thus combines high accuracy with high
efficiency. The price one has to pay is that the actual construction
of well localized WFs is not always straightforward, and requires some
user interaction - in particular for metallic systems. Also the lack
of finite support of the WFs is unwanted in the context of transport,
although in practice it is not a serious problem since the WFs are well
localized. Finally, as already explained above, the risk of obtaining
different WFs for two similar but non-identical systems renders it
less straightforward to patch the parts together using the
Hamiltonians obtained for the separate calculations.

Most of the disadvantages of the WF basis are resolved by the PAO
basis set: by construction they have finite support and are identical
as long as the atomic species on which they are located are the
same. This renders it straightforward to patch together Hamiltonians
for separate sub systems as long as the KS potential can be smoothly
matched at the interfaces. On the other hand, to obtain an accuracy
matching the WFs, one needs to use a significantly larger number of
orbitals and thus longer computation times as compared to the WF method.
  \end{subsection}
\end{section}

\begin{section}{Gold chain with CO}\label{sec:AuCO}
  In this section we calculate the conductance of an infinite gold
  chain with a single CO molecule adsorbed. Scanning tunneling
  microscope (STM) experiments suggest that CO strongly depresses the
  transport of electrons through gold wires~\cite{au_co_experiments}. This has been supported by NEGF-DFT
  calculations~\cite{au_co_transport} which shows that the transmission function indeed 
  drops to zero at the Fermi level. The use of infinite gold chains
  as leads is clearly an oversimplification of the real situation,
  however, the model seems to capture the essential physics, i.e. the
  suppression of the conductance, and furthermore is well suited as a
  benchmark system due to its simplicity.

  The geometry of the system is shown in Fig. \ref{fig.fig1ab}(a). We
  use a supercell with transverse dimensions $12 \text{\AA} \times 12
  \text{\AA}$ and take all bond lengths from
  %!!!check bond lengths in the paper!!!
  Ref.~\onlinecite{au_co_transport}: $d_{\text{Au-Au}}=2.9~\text{\AA}$,
  $d_{\text{Au-C}}=1.96~\text{\AA}$, and
  $d_{\text{C-O}}=1.15~\text{\AA}$. The Au atom holding the CO is
  shifted towards CO by $0.2~\text{\AA}$. In method 1 we obtain six WFs per Au
  atom and seven WFs for the CO molecular states. Due to
  the elongated bond length of the Au-wire, we found it necessary in method 2 to
  increase the range of the Au PAOs in order to converge the band-structure
  of the Au-wire.  The confinement energy was therefore in this case
  set to $10^{-4} \text{Ry}$.

  \begin{figure}[!h]
   \includegraphics[width=0.95\linewidth,angle=0]{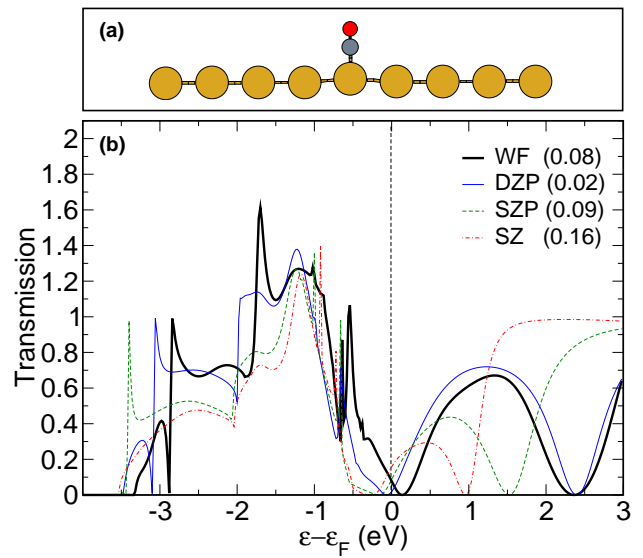}
   \caption[cap.Pt6]{\label{fig.fig1ab} (color online).
                     (a) Central region used to describe a single CO molecule adsorbed on a monatomic Au wire.
                     (b) Transmission functions for the Au wire CO system calculated using method 1 (WF) and method 2 for three different PAO basis sets.  
                      The transmission function at the Fermi level 
                       is indicated in the parenthesis following the legends.
                      }
  \end{figure}

  In Fig. \ref{fig.fig1ab}(b) we show the calculated transmission
  function using three different PAO basis sets and the WF basis set.
  Overall, the PAO result approaches the WF result as the basis set is
  enlarged. For the largest PAO basis (DZP) the agreement is in fact
  very satisfactory given the differences in the underlying DFT
  methods, e.g. ultrasoft- versus norm-conserving pseudo potentials. 
  The remaining difference can be further reduced by a rigid
  shift of the DZP transmission by about $0.15~\text{eV}$. 
  %This turns out to be a general trend to which we shall return later...

  All transmission functions feature an anti-resonance near the Fermi
  level. However, upon enlarging the PAO basis the position of the
  anti-resonance shifts as follows: (SZ) $-0.27~\text{eV}$, (SZP)
  $-0.16~\text{eV}$, (DZP) $-0.06~\text{eV}$, and (WF) $0.12~\text{eV}$.
  Note that the position of the anti-resonance obtained with the WFs
  is approached as the PAO basis set is increased. Also, the curvature of
  the anti-resonance is improved as the PAO basis set is enlarged. 
  The improvement in these features are, however, not directly reflected in the
  conductances indicated in the parenthesis following the legends 
  in Fig. \ref{fig.fig1ab}(b). The reason for the this apparent disagreement is
  the rigid shift between the PAO and WF transmission functions.
  
  We observe that our results differs from the calculation in 
  Ref.~\onlinecite{au_co_transport}: While the latter finds two peaks in
  the energy range $0-2~\text{eV}$ our converged transmission function
  shows a single broad peak. In general, both our PAO and WF based
  transmission functions present less structure than the transmission
  function reported in Ref.~\onlinecite{au_co_transport}. We suspect that
  these differences are 
  related to the way the coupling $V_{\alpha C}$ is calculated.
\end{section}

\begin{section}{Pt contact}\label{sec:pt_contact}
  Atomic point contacts formed from late transition metals such as Au,
  Pt, and Pd show very stable and reproducible features in
  conductance measurements~\cite{agrait_report}. This, together with
  the simplicity implied by their homogeneity, makes them ideal as
  benchmark systems for transport calculations. Here we consider the
  conductance of a pure Pt contact for which both experimental
  conductance data~\cite{smit02,sknielsen,untiedt04} as well as
  theoretical calculations~\cite{prb_strange_pt_co,prl_pt_osc,
    palacios_pt_contact} are available.

  Conductance histograms obtained from mechanically controlled break
  junction experiments on pure Pt samples show a peak near
  $1.5~G_0$, indicating that as a Pt contact is pulled
  structures with a conductance around $1.5~G_0$ are frequently
  formed.  NEGF-DFT calculations have shown that (zig-zag) monatomic Pt
  chains indeed have a conductance close to this
  value~\cite{sknielsen,prl_pt_osc,prb_strange_pt_co}. Moreover, the
  calculations predict an increasing conductance as the Pt chain is
  stretched and evolves from a zig-zag to a linear configuration. This effect has also been observed experimentally~\cite{sknielsen}. %!!!check ref!!!
  \begin{figure}[!h]
   \includegraphics[width=0.95\linewidth,angle=0]{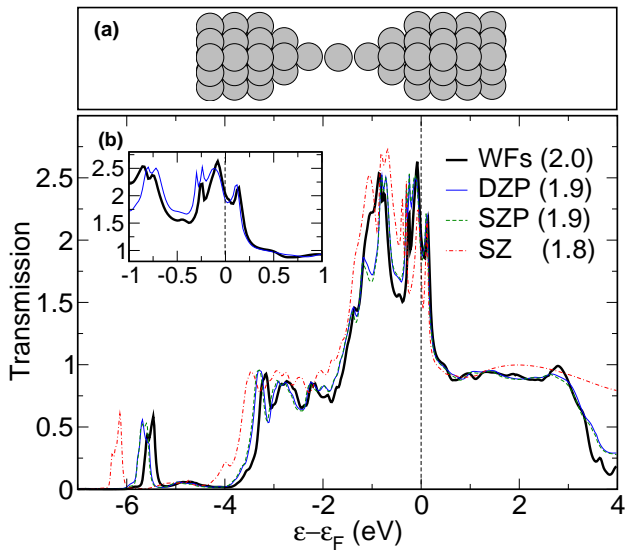}
   \caption[cap.Pt6]{\label{fig.fig2} (color online).
                     (a) Supercell used for the DFT calculation
                     of a short linear Pt chain between 
                     Pt(111) surfaces.   
                     (b) The calculated transmission function using
                     method 1 and method 2. The transmission at 
                     the Fermi level is indicated in the parenthesis following
                     the legends.
                     In the inset we show the transmission function
                     in the important region near the Fermi level 
                     for the DZP basis set and the WF basis set.
                     }
  \end{figure}

  In Fig. \ref{fig.fig2}(a) we show the supercell used to model the
  scattering region of the Pt contact. The Pt contact is modeled by
  two four-atom pyramids attached to (111) surfaces containing 3x3
  atoms in the surface plane. In order to ensure that the effective KS
  potential has converged to its bulk value at the end planes of the
  supercell we include 3-4 atomic layers (ABC-CABC stacking) on either
  side of the pyramids. The chain is formed by inserting a single Pt
  atom between the apex atoms of the two pyramids. We have relaxed the
  contact region (pyramids+chain) while keeping the rest of the
  structure fixed in the bulk configuration with lattice constant
  $3.93~\text{\AA}$ and a distance of $14.60~\text{\AA}$ between the (111) surfaces.
  The cutoff energy used in the construction of WFs was set to $\varepsilon_F+4.0~\text{eV}$ ensuring that the KS eigenstates below this value are exactly reproducible in terms of the WFs. 

  In Fig. \ref{fig.fig2}(b) we show the calculated transmission functions
  using method 1 and method 2. The qualitative agreement between 
  the two methods is striking, however, only the SZP and DZP basis sets
  provide quantitative agreement with the WF result.
  The SZ basis set results in a down shift of the peak at 
  $-6~\text{eV}$ and an incorrect description of the features in the important
  region near the Fermi level.
  Here the converged transmission function shows two peaks positioned at
  $\varepsilon_F-0.8~\text{eV}$ and just below the Fermi level, respectively. 
  The
  main peak astride the Fermi level in fact consists of three smaller peaks, as
  seen more clearly in the inset for the DZP and WF basis set.  
  These particular features in the transmission function were also
  observed in Ref.~\onlinecite{palacios_pt_contact} for a similar Pt contact, employing a method based on quantum chemistry
  software and a description of the bulk electrodes by a semi empirical
  tight-binding Hamiltonian on a Bethe lattice~\cite{palacios_transport_code}. 
  Also, the calculated conductance of
  $2.3~G_0$ is in agreement with our results, considering the
  structural differences.

  In Fig. \ref{fig.fig3} we show the calculated conductance of
  the Pt contact for three electrode displacements:
  $13.62~\text{\AA}$, $14.60~\text{\AA}$, and $14.75~\text{\AA}$.
  The three
  configurations correspond to an unstrained Pt chain, the chain just
  before it snaps, and the broken chain, respectively.

  \begin{figure}[!h]
   \includegraphics[width=0.95\linewidth,angle=0]{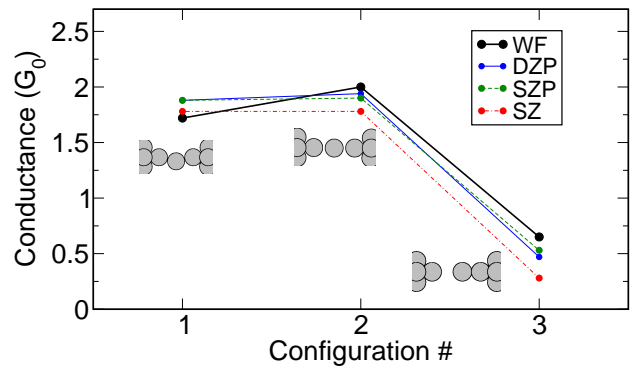}
   \caption[cap.Pt6]{\label{fig.fig3}(color online).
                     Conductance for three different
                     configurations during the stretching
                     of a small Pt chain. Configurations 1,2, and 3
                     correspond to the unstrained chain,
                     maximally strained chain, and a broken chain,
                     respectively. The contact atoms are shown in the insets}
  \end{figure}

  All basis sets, except for the SZ, are able to reproduce the trend of increasing
  conductance prior to rupture. The
  SZ basis set underestimates the absolute conductance by nearly $0.5~G_0$
  in the strained and broken configurations as compared to the WF result.
  The conductance calculated with the SZP and DZP basis set is almost
  identical and shows results more consistent with the WF basis for
  all three configurations.
 
\end{section}
\begin{section}{Pt-$\text{H}_2$-Pt contact}\label{sec:pt_h2}
  In this section we consider the simplest possible molecular
  junctions, namely a single hydrogen molecule between metallic Pt
  contacts. Like the metallic point contacts, the Pt-$\text{H}_2$-Pt
  junction shows stable and reproducible behavior in conductance
  measurements. In particular, a very pronounced peak close to $1G_0$
  appears in the conductance histogram obtained when a Pt contact is
  broken in a hydrogen atmosphere~\cite{smit02}. Although reported conductance calculations show significant variation (see below), there have been given substantial evidence
  that the structure responsible for the peak consists
  of a single hydrogen molecule bridging the 
  Pt contacts~\cite{smit02,djukic}.

  Several groups have published NEGF-DFT calculations for the
  transmission function of the Pt-$\text{H}_2$-Pt system. Most 
  find a conductance of $(0.9-1.0)G_0$, but also much lower values of
  $(0.2-0.5)G_0$ have been reported.
  \cite{smit02,prl_thygesen_h2,prb_ptpd_h2,cuevas,garcia}.

  \begin{figure}[!h]
  \includegraphics[width=0.95\linewidth,angle=0]{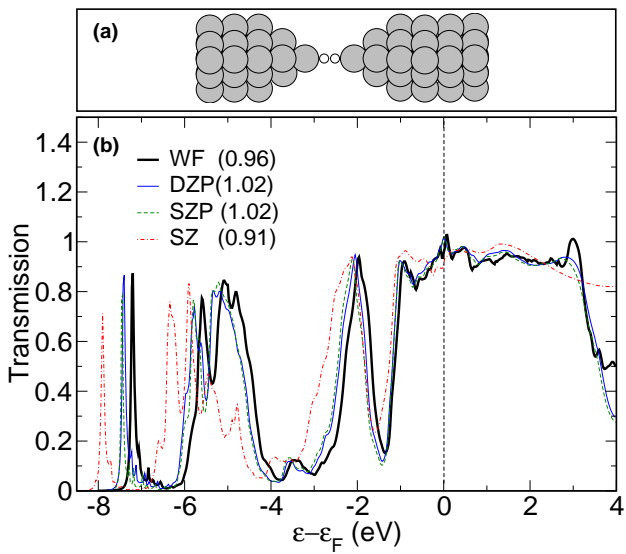}
  \caption[cap.Pt6]{\label{fig.fig4} (color online). (a) Supercell used to model the central region of the $\text{Pt}-\text{H}_2-\text{Pt}$ junction. (b) Transmission function for the Pt
    Hydrogen bridge. 
    The transmission function at the Fermi level 
    is indicated in the parenthesis following the legends.}
  \end{figure}

 For the benchmark calculations we use the same setup as in Sec. \ref{sec:pt_contact} with the
  central Pt atom replaced by a hydrogen molecule, see Fig. \ref{fig.fig4}(a).  The relevant bond length determining the
  structure after relaxation of the Pt pyramids and the hydrogen atoms
  are $d_{\text{Pt-H}}=1.7~\text{\AA}$ and
  $d_{\text{H-H}}=1.0~\text{\AA}$.

  In Fig. \ref{fig.fig4}(b) we show the calculated transmission
  functions.  Like in the case of the Pt contact, the agreement between the different calculations is striking,
  especially in the important region around the Fermi level.  The SZ
  basis set reproduces the qualitative features of the larger basis
  sets, but introduces a considerable down shift of the low-lying peaks.

  The very good agreement between the two methods indicates that the
  transmission function for this system is rather insensitive to the
  basis set. We stress, however, that a proper $\mathbf k_{\parallel}$-point sampling of the transmission
  function is crucial to obtain meaningful results independently of the quality of the basis set. Restricting the calculation to the $\Gamma$ point yields a
  transmission function with a (unphysical) peak at the Fermi
  level~\cite{prl_thygesen_h2}. We note in passing that such a peak is present in
  the transmission function reported in Ref.~\onlinecite{cuevas}. Such unphysical features resulting from an insufficient $\mathbf k_{\parallel}$-point sampling are not properties of the
  molecular junction, but are rather due to van Hove singularities in
  the quasi one-dimensional leads~\cite{prb_thygesen_kpoints}.  The
results reported in Ref.~\onlinecite{prb_ptpd_h2} are based on Siesta DFT code and show good agreement with our results. The
  conductance obtained in one of the early theoretical
  calculations~\cite{garcia} on the hydrogen molecular bridge are
  considerably lower than our and most other results. The
  calculational method applied in Ref.~\onlinecite{garcia} is,
  however, the same as applied in the study of pure Pt contacts~\cite{palacios_pt_contact}, which agrees well with our results as
  discussed in Sec.\ref{sec:pt_contact}. We speculate if this could be related to the smaller
  size of the Pt cluster used to model the electrodes in \onlinecite{garcia} as compared to \onlinecite{palacios_pt_contact}. Another possibility for the discrepancies
  is the use of the B3LYP energy functional in \onlinecite{garcia}
  instead of an LDA/GGA functional used in most other works.

\end{section}

\begin{section}{Benzene-1,4-dithiol (BDT) between Au(111) surfaces}\label{sec:AuBDT}
The Benzene-1,4-dithiol (BDT) molecule suspended between gold electrodes was among the first single-molecule junctions to be studied and has become the standard reference for calculations of nano-scale conductance. Depending on the experimental setup, measured conductances vary between $10^{-4}~G_0$ and $10^{-1}~G_0$~\cite{m_a_reed_bdt,tsutsui_bdt,tao_bdt,ulrich_bdt,ghosh_bdt}, while the calculated values typically lie in the range $(0.05-0.4)~G_0$~\cite{prb_evers_2004,prb_kondo_2006,cms_stokbro_2003,prb_faleev_2005,
        prl_lang_2000,prl_pantelides_2007,jcp_sankey_2004, prl_emberly_2003}.
In general it has been found that the transmission function is strongly
  dependent on the bonding site of the S atom~\cite{prb_kondo_2006,jcp_sankey_2004},
  while variations in the Au-S bond length only affects the transmission function weakly~\cite{prb_evers_2004}.

As our objective is to establish a computational benchmark for the Au-BDT system we choose the simple junction geometry shown in Fig. \ref{fig.AuBDT}(a). The S atoms are placed in fcc hollow sites of the Au(111) surface 
  and the molecule has been relaxed while keeping the Au atoms fixed in the bulk crystal structure. We use an Au lattice constant of $4.18~\text{\AA}$, and a distance between the Au(111) surfaces of $9.68~\text{\AA}$. With these constrains the relevant bond lengths are: 
  $d_\text{Au-S}\text{=2.45 \AA}$, $d_\text{S-C}\text{=1.73 \AA}$,
  and $d_\text{C-H}\text{=1.09 \AA}$.

  \begin{figure}[!h]
  \includegraphics[width=0.95\linewidth,angle=0]{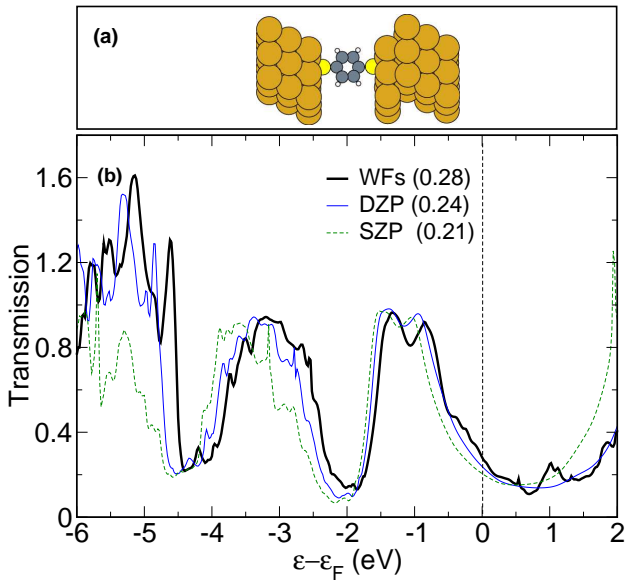}
  \caption[cap.Pt6]{\label{fig.AuBDT} (color online).  (a) Supercell used to model the central region of the Au(111)-BDT-Au(111) system with S at the fcc hollow site.  
    (b) The calculated transmission functions. Note, that the SZ transmission
    function has been omitted for clarity. The transmission function at the Fermi level
         is indicated in the parenthesis following the legends.}
  \end{figure}

  In Fig. \ref{fig.AuBDT}(b) we show the calculated transmission functions (the SZ result has been omitted for clarity). Notice that we plot the transmission function only up to $2~\text{eV}$ above the Fermi level. This is because the the WF transmission at larger energies is sensitive to the parameters used in the construction of the WFs, in particular the cutoff energy $E_0$, and thus we cannot be sure about the WF result above $2~\text{eV}+\varepsilon_F$.

  The three transmission functions agree very well in the energy range 
  from $2~\text{eV}$ below the Fermi level to $1~\text{eV}$ above the 
  Fermi level, while only the DZP result agrees quantitatively with the WF result in the entire energy range. We again notice the down shift of the PAO transmission functions relative to the WF result.
  
  The presence of a broad transmission peak positioned $\sim 1~\text{eV}$ below the Fermi level is in qualitative agreement with previous
  results~\cite{prb_kondo_2006,prb_evers_2004,prb_faleev_2005,cms_stokbro_2003,jcp_xue_2001,
basis}. For more stretched configurations, i.e. for larger values of the S-C bond length, than the one used in the present study, the broad peak splits into two more narrow peaks~\cite{jcp_thygesen_transport}.
  
The transmission function presented in 
  Ref.~\onlinecite{cms_stokbro_2003} was obtained
  using a method very similar to our method 2, however, the reported conductance of $0.4~G_0$ is
  almost twice as high as our DZP results of $0.24~G_0$.
  The large conductance arises because the transmission peak
  closest to the Fermi level is considerably broader than what we find.
  If, however, we restrict the $\mathbf k_{\parallel}$ to the $\Gamma$-point 
  we find the same broadening as in Ref.~\onlinecite{cms_stokbro_2003}
  and a very similar conductance of $0.37~G_0$. Another feature of the $\Gamma$-point only transmission function is that the second peak positioned at $\sim 3~\text{eV}$ below the Fermi level separates into a number of more narrow peaks.
   
  In Ref.~\onlinecite{prb_faleev_2005} the transmission function
  is calculated from the LMTO-ASA method and averaged over 
  36 irreducible $\mathbf k_{\parallel}$-points. Both the width and the position of the two
  peaks in the transmission function at $1~\text{eV}$ and $3~\text{eV}$ 
  below the Fermi level, are in good agreement with our results.
  The height of the former peak is, however, lower than in our calculation and this reduces the
  conductance to a value of $0.07~G_0$. We suspect that this difference could be due to differences in the adopted contact geometries. 
 
\end{section}

\begin{section}{Bipyridine between Au(111) surfaces}
As the last reference system we consider a bipyridine molecule attached between
two gold-electrodes. STM experiments on bipyridine molecules in a toluene 
solution~\cite{tao} 
show that the conductance of Au-bipyridine junctions is quantized in multiples of $0.01 G_0$ which was interpreted as the
formation of stable
contacts containing one or more molecules. The conductance
is expected to be sensitive to the details of the contact
geometry~\cite{prb_stadler_2005}, however, for the benchmark calculation we
choose the simple case of binding at the on-top site of a flat Au(111) surface,
as shown in Fig. \ref{fig.fig5}(a). The Au electrodes are the same as
used for the BDT molecule in Sec.\ref{sec:AuBDT}, the Au(111)-N distance 
is $2.180~\text{\AA}$, while the electrode displacement is $11.53~\text{\AA}$.

The transmission functions calculated using either PAOs or WFs
are shown in Fig. \ref{fig.fig5}(b). At first it is noted that the overall structures of the transmission functions are very
similar. In the Siesta calculations, the position of the narrow LUMO peak which governs the transport is underestimated but converges towards the WF result as
the PAO basis set is enlarged, see the inset of 
Fig. \ref{fig.fig5}(b). The alignment of the LUMO energy level with respect to
the Fermi level and its relation to charge transfer was studied in
Ref.~\onlinecite{prb_stadler_2006}. 

Several groups have investigated the transport properties of bipyridine-gold
junctions, and there is general agreement
that the low bias conductance depends crucially on the details of the contact 
geometry. As different groups have chosen different geometries and models for
the gold electrodes a direct comparison of the reported transmission functions is difficult.

To the best of our knowledge the first theoretical paper on the bipyridine system
is by Tada \emph{et al.}~\cite{tada}. In their calculations bipyridine is adsorbed on-top of an Au-atom of a rather small Au cluster, and the coupling to the infinite electrodes is modeled by a broadening parameter fitted to experimental
data. The zero-voltage transmission function contains
some of the same peak structures as we observe.
Hou \emph{et al.}~\cite{hou} have published several papers on the gold-bipyridine
juncition. Like Tada \emph{et al.} they include only a few gold atoms in the \emph{ab-initio} calculation and treat the coupling to electrodes through
a model self-energy term. The peak structure of the transmission function is quite different from ours. This could be due to the small size of the gold clusters used to mimick the electrodes.
While most other groups observe tunneling through the LUMO tail~\cite{perez,prb_stadler_2005,prb_stadler_2006}, Hou \emph{et al.} argue that
the transport is mainly taking place via the HOMO-2 state. 
Calculations by Wu \emph{et al.}~\cite{wu1,wu2} obtained using a Siesta-based transport code~\cite{transiesta}, for bipyridine attached to the on-top site of a gold surface show overall good agreement with our results (see Fig. 7(a) in paper \onlinecite{wu1}). The minor 
differences are probably related to the fact that only the $\Gamma$-point has been used in the latter paper.

  \begin{figure}[!h]
   \includegraphics[width=0.95\linewidth,angle=0]{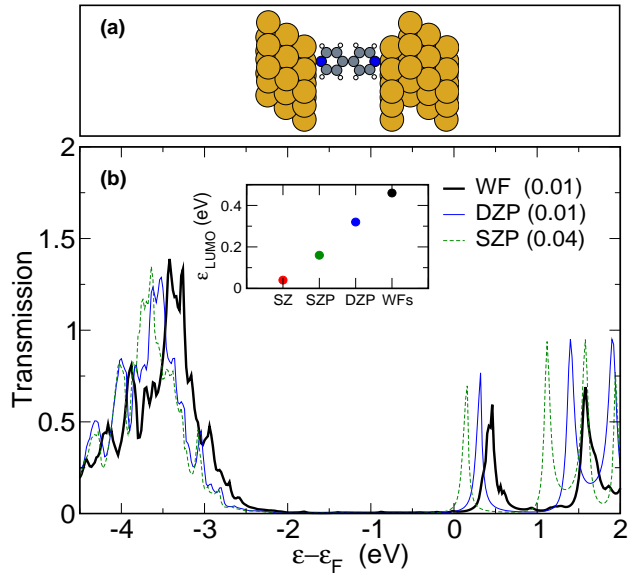}
   \caption[cap.Pt6]{\label{fig.fig5} (color online). (a) Supercell used to describe the central region of the bipyridine-Au(111) junction. (b) Calculated transmission
     functions (the SZ result has been omitted for clarity). The inset shows the dependence of the LUMO position on the basis sets. The transmission function at the Fermi level 
     is indicated in the parenthesis following the legends.}
  \end{figure}

\end{section}

\begin{section}{Conclusions}\label{sec:conclusions}
  We have established a set of benchmark calculations for the Kohn-Sham(PBE) elastic transmission function of five representative single-molecule junctions using two different methods based on independent DFT codes: (i) A
  plane wave DFT code in combination with maximally localized Wannier functions. (ii) The Siesta program which applies finite range pseudoatomic orbitals.

For all five systems we find that the Siesta result converges towards the WF result as the Siesta basis is enlarged from SZ to DZP with the latter yielding very good quantitative agreement with the WF transmission. In the Siesta calculations the transmission peaks relative to the peaks obtained with the plane-wave calculation are systematically shifted toward lower energies. The problem can be overcome by enlarging the Siesta basis, however, the convergence can be rather slow. 

\end{section}
\begin{section}{Acknowledgments}
  The authors acknowledge support from the Danish Center for
  Scientific Computing through grant HDW-1103-06. The Center for
  Atomic-scale Materials Design is sponsored by the Lundbeck
  Foundation.
\end{section}
\newpage

%%%%%%% References

%\bibliographystyle{apsrev}
\bibliographystyle{unsrt}

\end{document}